\documentclass[iop,revtex4]{emulateapj}
\usepackage{lineno}

\begin{document}

\renewcommand{\topfraction}{1.0}
\renewcommand{\bottomfraction}{1.0}
\renewcommand{\textfraction}{0.0}

\title{New orbits based on speckle interferometry at SOAR. II.$^1$}

\altaffiltext{1}{Based on observations obtained  at the Southern Astrophysical Research
(SOAR) telescope. }

\author{Andrei Tokovinin}
\affil{Cerro Tololo Inter-American Observatory, Casilla 603, La Serena, Chile}
\email{atokovinin@ctio.noao.edu}

\begin{abstract}
Orbits of  44 close  and fast visual  binaries are computed  using the
latest speckle  observations; 23 orbits  are determined for  the first
time,  the rest are  revisions, some  quite substantial.  Six combined
orbits use  radial velocities.  The  median period is 15.6  years, the
shortest  period  is  one   year.  Most  stars  are  nearby  late-type
dwarfs. Dynamical  parallaxes and estimates of the  masses are derived
from the orbital elements and the photometry of the components.
\end{abstract} 

\maketitle

\section{Introduction}
\label{sec:intro}

This  paper presents new  or updated  orbits of  44 binary  systems or
subsystems. It  is based on speckle  interferometric measurements made
at  the  4.1  m   Southern  Astrophyisical  Research  (SOAR)  telescope
\citep{TMH10,SAM09,Tok2012b,Tok2014a,Tok2015c,SAM2015}  combined  with
archival  data collected in  the Washington  Double Star  Catalog, WDS
\citep{WDS}.  It  continues previous  work on binary  orbits resulting
from the SOAR speckle program  and follows the template of the Paper~I
\citep{Tok2016},   where  the   motivation  is   discussed.   Briefly,
calculation   of   binary  orbits   is   part   of  the   astronomical
infrastructure and visual orbital elements are used in many areas. The
state of  the art   is reflected in the  Sixth Catalog of
Visual       Binary      Orbits,       VB6      \citep{VB6}.\footnote{
  \url{http://www.usno.navy.mil/USNO/astrometry/optical-IR-prod/wds/orb6.html}.}

Accurate parallaxes from {\it Gaia} \citep{Gaia} are expected to yield
good   mass   measurements    of   visual   binaries   with   reliable
orbits. However, the  short 5 year duration of  the {\it Gaia} mission
will make it difficult  to disentangle orbital and parallactic motions
of  the photo-center  and might  lead to  biased parallaxes  of visual
binaries,   as    happened   with   the    {\it   Hipparcos}   mission
 \citep{ST1998,S99}.  Therefore,  orbital elements based  on ground-based
data will be essential for  getting correct astrometry of binary stars
from {\it  Gaia}.  The {\it  Gaia} DR1 explicitly avoids  binaries and
gives parallaxes only for a few stars featured in this paper.

\section{Orbital Elements}
\label{sec:orb}

\begin{figure*}
\epsscale{1.0}
\plotone{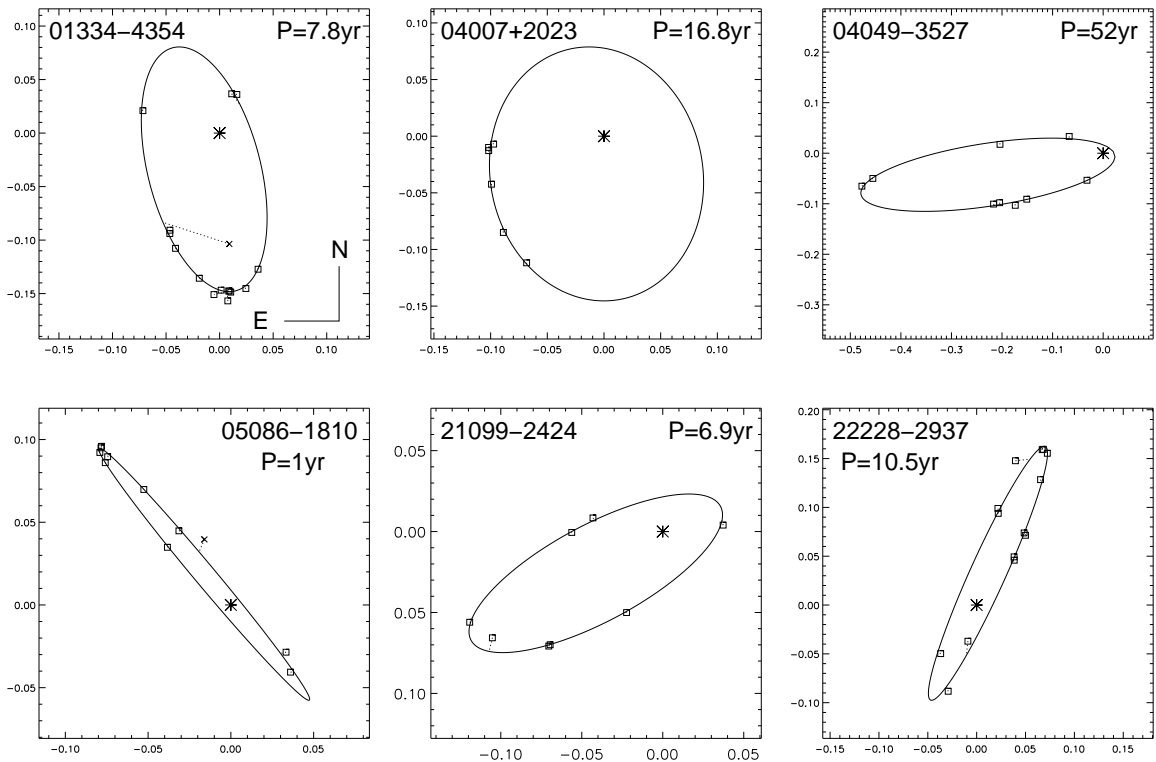}
\caption{New  reliable  orbits of  six  binaries.  In  each plot,  the
  primary component (star) is at  the coordinate origin, with the scale in
  arcseconds.  The ellipse depicts the orbit, the squares connected to
  the ellipse  are the measurements. Measurements with  large deviations and
  low weight are plotted as crosses.
\label{fig:orb}  }
\end{figure*}

The methods  are identical to those  of Paper I, so  they are outlined
only  briefly.  Orbital elements  and their  errors are  determined by
least-squares  fitting  using  the  IDL  code  {\tt  ORBIT}.\footnote{
  \url{http://www.ctio.noao.edu/\~{}atokovin/orbit/index.html}  } When
no  prior orbit  is  available, the  initial  approximation is  chosen
interactively, considering the observed motion of each pair. This step
is largely heuristic.   Sometimes the same data can  be represented by
several  very  different  orbits.     In  those  cases,  additional
  information such as the mass  sum, radial velocities (RVs), times of
  non-resolutions, etc.  is used  to select the correct orbit; several
  examples are found in Section~\ref{sec:notes}. 

The  second step  is  the least-squares  refinement  (fitting) of  the
elements. The weights are inversely  proportional to the square of the
measurement  errors,  which are  assigned  subjectively  based on  the
technique  used. Some outliers  are given  very low  weight, otherwise
they would distort the orbit.  For pairs observed at SOAR, the typical
position errors  are 2\,mas in  the $y$ filter  and 5\,mas in  the $I$
filter. They are larger for  difficult binaries which are either close
to the  diffraction limit or very  faint.  When the orbit  is not well
constrained by the data, some orbital { element, e.g.  eccentricity
  or inclination, is fixed to  the value that results in the expected
mass sum,  given the HIP2  parallax \citep{HIP2}.   The  choice of
  the element to fix is  individual, depending on the coverage and the
  orbit configuration.  Its value is selected by trial and error while
  fitting  the   remaining  elements   and  checking  the   mass  sum.
  Considering  potential  parallax  biases  mentioned  above  and  the
  uncertainty of the estimated  masses, this approach is hazardous; it
  is  used only  for  preliminary low-quality  orbits.     RVs,  when
available from  the literature, are  used jointly with  the positional
measurements,  improving substantially  the quality  of  such combined
orbits.

Table~\ref{tab:orb}  lists the  orbital elements  and their  errors in
common notation ($P$  -- orbital period, $T_0$ --  epoch of periastron
 in Besselian years, $e$  -- eccentricity, $a$ -- semimajor axis,
$\Omega$  -- position  angle  of  the node  for  the equinox  J2000.0,
$\omega$ --  argument of periastron,  $i$ -- inclination).   The first
column gives  the WDS code of  the binary and, in  the following line,
its {\it Hipparcos} number.  The  system identifier adopted in the WDS
(``discoverer code'')  is given in  the second column. For  each pair,
the first line contains the orbital elements, while the following line
gives  their formal  errors.   The orbit grades  are given  in  the VB6
  system, where grades 4 and 5 mean preliminary orbits and grade 1 are
  definitive and accurate orbits. Those grades were kindly computed by
  the Referee. However,  this grading system does not  account for the
  RV    data    that    strengthen    combined    visual-spectroscopic
  orbits. Alternative  unofficial grades are provided  for such orbits
  in brackets.  The last column contains references to the previously
computed   visual   and    spectroscopic   orbits,   when   available.
Figure~\ref{fig:orb} gives  the plots of six new  but already reliable
orbits.

Individual    observations    and     residuals    are    listed    in
Table~\ref{tab:obs},  available in  full  electronically. It  contains
still unpublished measures  made at SOAR in 2016  and 2017, while some
published SOAR measures were  reprocessed. Its first column identifies
the binary by  its WDS code  (for  multiple systems, the discovery
  codes and component designations are found in Table~1). Then follow
the time  of observation  $T$  in  Besselian years,  the position
angle $\theta$ for the  time of observation (correction for precession
is done internally during orbit calculation), the separation $\rho$ in
arcseconds, and the measurement error $\sigma$.  Unrealistically large
errors are  assigned to the  discarded observations  to  give them
  zero weight, but those data are  still kept in the Table.  The {\bf
  following} two columns  of Table~\ref{tab:obs} contain the residuals
O$-$C in  angle and separation.  The last column  contains  flags
  specifying  the  source  of  the  data: M  for  historic  micrometer
  measures,  H  for  the   {\it  Hipparcos}  measure,  S  for  speckle
  interferometry  at SOAR,  I  for speckle  interferometry from  other
  sources  found  in  the  online  Fourth  Catalog  of  Interferometric
  Measurements of Binary Stars \citep{INT4}.\footnote{ \url{http://www.usno.navy.mil/USNO/astrometry/optical-IR-prod/wds/int4/fourth-catalog-of-interferometric-measurements-of-binary-stars}}

Table~\ref{tab:ptm}  provides additional  information, namely  the the
spectral  type as  given in  {\it Hipparcos}  or SIMBAD  and  the HIP2
parallax $\pi_{\rm  HIP2}$, to be  compared to the  dynamical parallax
$\pi_{\rm dyn}$  in the  next column. The  latter is evaluated  by the
Baize-Romani method, as explained  in Paper I.  The components' masses
${\cal M}_1$ and ${\cal M}_2$ obtained in this procedure are listed in
the columns  (6) and (7)  of Table~\ref{tab:ptm}.  Asterisks  mark the
dynamical  parallaxes  derived from  reliable  orbits  of  grade 3  or
better.   

The combined  $V$ magnitude and $V  - I_C$ color index  in columns (8)
and  (9) are  taken  mostly  from the  {\it  Hipparcos} catalog.   For
HIP~32366B, the combined magnitudes  refer to the secondary subsystem,
while $V -  I_C = 2.7$ mag is assumed to  match the estimated spectral
type.    The  last   four  columns   of   Table~\ref{tab:ptm}  provide
differential    photometry   resulting    from   the    SOAR   speckle
interferometry,  where  the  filters  $y$  and $I$  have  the  central
wavelengths and the bandwidths of 543/22 and 788/132 nm, respectively.
The   $\Delta  y$   and   $\Delta  I$   are   average  values,   while
$\sigma_{\Delta y} $ and $\sigma_{\Delta I}$ stand for the rms scatter
of  magnitude difference  in each  filter if  measured  several times,
indicating    the   internal    consistency   of    the   differential
photometry. For some pairs this scatter  is as good as 0.1 mag, but it
can be much larger.  The speckle differential photometry is unreliable
for very close  pairs at or below the diffraction  limit. For faint or
wide  binaries,  $\Delta   m$  can  be  systematically  overestimated.
Despite  these caveats,  speckle interferometry  at SOAR  is  the only
source of differential photometry for some close binaries.

\begin{figure}
\epsscale{1.0}
\plotone{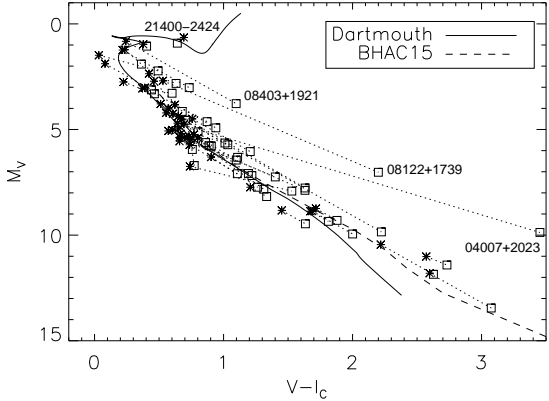}
\caption{Color-magnitude diagram. The primary and secondary components
  of each pair are plotted as asterisks and squares, respectively, and
  connected by dotted lines. The  full line is the Dartmouth isochrone
  \citep{Dotter2008},  the dashed  line is  the  BHAC15 \citep{BHAC15}
  isochrone, both for 1~Gyr age and solar metallicity. Four secondary
  components located above the main sequence are labeled by their WDS codes.
\label{fig:cmd}  }
\end{figure}

When  both $\Delta y$  and $\Delta  I$ are  measured, the   crude
empirical linear relation  $\Delta I \approx 0.7 \Delta  y$ holds 
  for dwarf stars observed at SOAR, allowing to estimate differential
magnitudes in both  filters even when measurements in  only one filter
are available.  It is also reasonable to assume that $\Delta V \approx
\Delta y$ and $\Delta I_C \approx \Delta I$. Therefore, magnitudes and
colors of  each binary component  can be estimated using  combined and
differential photometry.

Figure~\ref{fig:cmd} shows the  $(M_V, V-I_C)$ color-magnitude diagram
(CMD)  where  only  binaries   with  $\pi_{\rm  HIP2}  >  4$\,mas  are
plotted. For  reference, the 1~Gyr  isochrones from \citet{Dotter2008}
and \citet{BHAC15} are shown.

\section{Notes on Individual Binaries}
\label{sec:notes}

Brief comments on all binaries are provided in this Section. 

{\it 00135$-$3650.   } This is a  nearby pair of M0.5V  dwarfs.
With only four measures,  the orbit is obviously preliminary, although
this  system has  made  two  revolution since  its  discovery by  {\it
  Hipparcos}.  

{\it 00164$-$7024. } This first preliminary orbit lacks coverage.

{\it  00258+1025.}  The pair  of G-type  dwarfs HDS~57  approaches the
periastron, so fast motion is expected in the coming years.

{\it  01334$-$4354.} This is  the first,  but well-defined  orbit with
$P=7.77$\,years (Figure~\ref{fig:orb}).  The SOAR data sample one full
period.   The {\it  Hipparcos} measure  is  off because  of the  short
period.  The primary  mass of 1 ${\cal M}_\odot$ is  too large for the
spectral  type K0.5V  listed in  SIMBAD.   Both  components are
located  on the  main  sequence in the CMD. 

{\it 02166$-$5026.}   The pair TOK~185,  discovered at SOAR in  2011, has
covered more than a half of its 14 year orbit, including the
periastron; the orbit is updated.

{\it 02390$-$5811.}   This very fast ($P=6$\,years) pair of M0V dwarfs was
near the periastron in 2016.9 and has made four revolutions since its
discovery by {\it Hipparcos}.

{\it 04007+2023.}  This  star is a member of the  Hyades cluster and a
single-lined  spectroscopic  binary   with  a  period  of  16.7\,years
\citep{Griffin2012}.   The SOAR  measures  cover only  a  part of  the
ellipse,  but the  combined  orbit  using the  Griffin's  RVs is  well
defined  (Figure~\ref{fig:orb}).  Griffin  remarked  on the  potential
resolution of  this pair:  ``The difference in  luminosity in  the $V$
band may  be expected  to be around  five magnitudes --  an impossible
prospect visually...''  In fact the binary was resolved at SOAR a year
after the  publication of  the Griffin's paper.   Indeed, $\Delta  y =
5.2$ mag, but the pair is far from being ``impossible''. The secondary
component  is located  well  above the  main  sequence in  the CMD  in
Figure~\ref{fig:cmd}, having $V - I_C  = 3.45$ mag.  However, the five
measurements of  $\Delta I$ have a  large scatter of 0.73  mag, so the
unusually  red color  of the  secondary component  needs confirmation.
The  RV amplitude  corresponds to  the secondary  mass of  0.58 ${\cal
  M}_\odot$ if the primary mass is 1.14 ${\cal M}_\odot$, in 
agreement with the masses  in Table~\ref{tab:ptm}.

{\it 04049$-$3527.  }  This is a visual triple  system where the outer
pair A,BC is the 1\arcsec ~binary I~152 with the estimated period of
$\sim$300\,years.   Although the  52  year  orbit of  the  subsystem B,C  is
computed   for    the   first    time,   it   is    well   constrained
(Figure~\ref{fig:orb}); this pair,  discovered in 1990.9 at 0\farcs48,
passed through  the periastron in 2009.7  and now opens  up again. Its
large eccentricity of 0.92 could  be caused by the outer system.

{\it 04074$-$6413. }   The spectroscopic elements from \citet{Jen2015}
are fixed and only the  period and the ''visual'' elements are fitted.
Unfortunately,  no RVs  are  given  in that  paper  for computing  the
combined orbit.   The spectroscopic  mass of the  secondary component,
0.53 ${\cal M}_\odot$, matches the mass in Table~\ref{tab:ptm}.

{\it 04107$-$0452.}   This is  HIP~19508, HD~26441, or  ADS~3041.  The
previous  20 year  visual  orbit  with high  eccentricity  is that  of
\citet{Tok2014a}. A concordant  double-lined spectroscopic orbit based
on 36 years of RV coverage  has been published by \citet{Gri2015} in a
paper which is, unfortunately, ignored  by SIMBAD. The orbit here is a
combined one based on the Griffin's RVs and the speckle data, with the
visual  micrometer measures  given a  very low  weight.   The combined
orbit is highly  accurate and deserves the grade  1; its spectroscopic
elements ($K_1 = 11.61$, $K_2 = 12.55$, $V_0 = 26.59$ km~s$^{-1}$) are
close  to those given  by Griffin.   The combined  orbit leads  to the
masses of $1.120 \pm 0.017$ and $1.036 \pm 0.016$ ${\cal M}_\odot$ for
the  primary  and  secondary  components, respectively.   The  orbital
parallax of 17.22$\pm$0.31 mas is more accurate than the HIP2 parallax
16.09$\pm$0.65  mas.  The  stars  are slightly  evolved  off the  main
sequence,  hence  the   masses  in  Table~\ref{tab:ptm}  are  somewhat
over-estimated.  After submission of this paper we became aware of the
work  by  \citet{Docobo2017}  who  proposed a  similar  visual  orbit.
However, they have not computed the combined orbit and apparently used
a  different  weighting  scheme.   Their orbit  is  consequently  less
accurate.  While the  combined orbit gives $a =  0\farcs167 \pm 0.002$
and $i =  69\fdg0 \pm 0\fdg5$, the  orbit by Docobo et al.  gives $a =
0\farcs161 \pm  0.002$ and  $i = 66\fdg4  \pm 0\fdg5$;  the difference
exceeds the formal errors.

{\it 04422+0259.}   The updated 58 year orbit of  A~2424 is now  well defined by
the  speckle data  alone; only  the first  visual measure  is  used to
constrain the period.

{\it 05086$-$1810.}  The first orbit of WSI~72 with a remarkably short
(for a  resolved binary) period of one  year (Figure~\ref{fig:orb}) is
well defined by the 11  speckle measures, leaving the rms residuals of
only  1.4 mas  (the latest  2016.96 measure  at the  diffraction limit
deviates more, so it was  given a low weight).  The dynamical parallax
of  118.5\,mas  appears  more  accurate  than  the  HIP2  parallax  of
108\,mas,  obviously biased  by  the photocenter  motion. The  deduced
masses of 0.3 ${\cal M}_\odot$ are normal for the M5V dwarfs.

{\it 05103$-$0736.}  The nearly  circular 37.6 year orbit of HIP~24076
by \citet{Hrt2012a} fits the measures well, although it predicts a too
small mass  sum of 1.2 ${\cal  M}_\odot$.  Double lines  were noted by
\citet{N04}.  Spectroscopic  monitoring by N.   Gorynya (2017, private
communication) did not show any substantial velocity variability during
four years,  meaning that  the system is  not triple.   Therefore, the
orbit of the visual binary  is likely eccentric, with half the period.
The new  18.9 year orbit  with $e=0.82$ fits the  interferometric data
very  well (rms  residuals 2\,mas).   It yields  the mass  sum  of 2.7
${\cal M}_\odot$, appropriate for a couple of F8V stars.  Observations
around  the next  periastron in  2018.6  are expected  to confirm  the
eccentric orbit and  to refute the circular one.   The RV measurements
near the periastron can yield accurate masses.

{\it  05525$-$0217.}  The  12 year   orbit  of  HDS~787,  updated  from
\citep{Tok2014a},  is  now definitive,  with  a  good coverage.   Four
speckle  measures made  at  the  3.5 m  WYIN telescope  in  1999--2004  were
assigned low weights,  although the reason why they  are discrepant is
not  clear.  The  {\it Hipparcos}  magnitude difference  $\Delta  Hp =
0.87$ mag is  underestimated (in fact $\Delta y =  1.6$ mag), which is
normal  for such a  close (97\,mas) pair,  well below  the diffraction
limit of the 30 cm {\it Hipparcos} aperture.

{\it 06454$-$3148.} This is a  nearby triple system GJ~245.1 where the
faint pair  of low-mass  dwarfs Ba,Bb at  1\farcs4 from the  main star
HIP~32366 has  been discovered in 2005 by  \citet{Ehr2010}. Apart from
the discovery paper, the only published observations are those made at
SOAR.  The magnitude difference between the F7V primary and the late-M
component Ba is $\Delta I = 6.57  \pm 0.47$ mag, meaning that the faint pair Ba,Bb is
just above the  detection limit at SOAR. The 6.9  year orbit of Ba,Bb
is slightly  updated with respect to \citep{Tok2015c}.   With the {\it
  Gaia} parallax of 39.61\,mas it  corresponds to the mass sum of 0.54
${\cal M}_\odot$.   The mass of Ba  is therefore close  to 0.30 ${\cal
  M}_\odot$ and implies  an M3.5V dwarf.  Its magnitude  $I =11.9$ mag
according to our photometry roughly matches the standard relations and
the  distance modulus  of 2.0  mag.  Given  the lack  of  the $V$-band
photometry,  Table~\ref{tab:ptm}  adopts  the combined  magnitude  and
color  appropriate  for  the  masses, making  the  dynamical  parallax
meaningless.

{\it 06533$-$1902.  }  The first 36 year orbit of  CHR~169 is reasonably
well constrained, despite  the large gap in its coverage  between 1996 and
2014. The pair is actually going through the periastron.

{\it   07269+2015.}    The  orbit   of   CHR~26  by   \citet{Ole1998c}
 with $P=14.17$\,years did not match the latest measures; it is revised to
$P=8.6$\,years.

{\it  08122+1739.   }   This  is  $\zeta$  Cnc  C,  a  member  of  the
hierarchical multiple system. The 59  year visual orbit of A,B and the
crude  visual orbit  of AB,C  (period 1115  years) have  been computed
previously.    Here  the   preliminary  visual   orbit  of   Ca,Cb  by
\citet{RAO2015} is  replaced by the  combined 17 year orbit  using the
RVs  from  \citet{Griffin2000}.  Our  orbit  agrees  roughly with  the
astrometric  orbit computed  by \citet{Heintz1996}.   However,  only a
small  sector of  the visual  orbit is  covered by  accurate measures,
whereas the  remaining measures  have large errors  and fit  the orbit
poorly. The HIP2 parallax of 39.9\,mas  leads to the large mass sum of
2.45    ${\cal   M}_\odot$    and   supports    the    conclusion   of
\citet{Hutchings2000}  that the masses  of Ca  and Cb  are comparable,
with Cb being  a close pair of M2 dwarfs.  The  unusually red color of
Cb  ($\Delta y =  4.3$, $\Delta  I =2.7$,  $\Delta K  = 0.2$  mag, see
Figure~\ref{fig:cmd}) is explained by  its binarity.  To reconcile the
RV amplitude with the mass  sum, we fixed the inclination to 150\degr.
The RV amplitude and the  inclination in the combined orbit correspond
to the  Cb mass  of 1.25  ${\cal M}_\odot$ if  the mass  of Ca  is 1.2
${\cal M}_\odot$.  The dynamical  masses in Table~\ref{tab:ptm} do not
account for the  binarity of Cb and are  therefore incorrect.  Further
speckle monitoring of this interesting pair is obviously needed.

{\it 08403+1921.} This is an A9V triple system in the Praesepe cluster
(NGC  2635),   where  the  visual   secondary  B  is   a  double-lined
spectroscopic binary with  a period of 48.7 days.   The outer orbit by
\citet{Ole2002b}  with $P=13.2$\,years corresponds  to the  very large
mass  sum of 54  ${\cal M}_\odot$,  using the  {\it Gaia}  parallax of
4.71\,mas; it does not match  the latest measures.  The available data
can  be represented by  a very  eccentric ($e=0.84$)  orbit with  $P =
10.6$ years or by a nearly  circular orbit with $P = 21.6$ years, both
with  the inclination of  90\degr.  However,  neither of  those orbits
match  the  separation  of  115\,mas  measured  in  1982.25  by  lunar
occultations. Therefore we prefer the third orbit with a longer period
of 35.5  years, compatible with  the occultation measure,  although it
fits  the speckle  data  slightly worse  than  the two  shorter-period
orbits.  The  mass sum in  the 35 year  orbit is reduced to  21 ${\cal
  M}_\odot$ with the {\it Gaia} parallax or to 6 ${\cal M}_\odot$ with
the HIP2 parallax  of 7.15\,mas. The HIP2 parallax  places the primary
component  on the main  sequence turnoff  (Figure~\ref{fig:cmd}).  The
problem with  this binary  is the lack  of speckle  monitoring between
1997 and 2016; during this period, only one measure was made.

\citet{Abt1999}  measured   the  RVs  of   the  primary  broad-lined
component  which show a  positive trend  during the  two year time span of
their data.  The trend is compatible  with all visual  orbits and does
not help to choose between them. The  RVs of the center of mass of the
secondary computed from their 48  day spectroscopic orbit have a large
scatter  and  throw  some  doubt  on that  orbit,  which  the  authors
themselves call tentative.

{\it  08447$-$4238.}    This  is  a  very  well   covered  orbit  with
$P=2.26$\,years. The HIP2 parallax is likely biased by the short period.

\begin{figure}[ht]
\epsscale{1.0}
\plotone{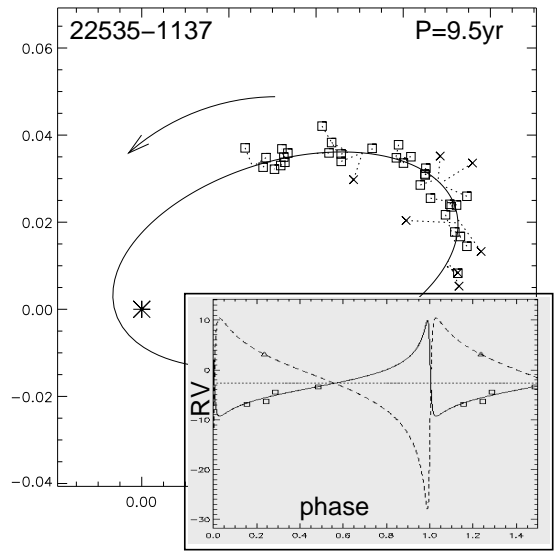}
\caption{The  eccentric  orbit of  HR  8704  with $P=9.5$\,years.  The
  insert shows the RV curve.
\label{fig:MCA73}  }
\end{figure}

{\it 09252$-$1258.}   The 27 year orbit of  WSI~73 by \citet{Tok2015c}
is radically revised  here to $P=13$ years after this  pair of K1V dwarfs
became unresolved at SOAR in 2015 and opened up in 2016. It will
close again in 2019. The new  orbit  appears secure.

{\it 10121$-$0241.} DEL~3 is a pair of nearby M0V dwarfs for which
we propose the first, still preliminary edge-on orbit with a short 6.4 year period. 
The eccentricity is fixed to tune the mass sum. 

{\it  10294+1211.}   The  orbit  by  \citet{Cve2016b}  ($P=23.36$  year)
strongly  disagrees  with  the   data  and  is  radically  revised  to
$P=15.6$\,years.  Although the pair has made nearly two revolutions since
its discovery in 1991, the coverage remains scarce.

{\it  10529$-$1717.}  The  updated   15 year  orbit  of  HDS~1556 
is now definitive, with both extremities covered. The
pair is closing down and will pass through the periastron in 2018.55.

{\it 11420$-$1701.} This visual triple  system was discovered at
SOAR in 2013.  The first preliminary 21 year orbit of  the inner pair TOK~384 is
proposed here (obviously, it  still lacks  coverage), while  the estimated
period of the outer 1\farcs1 companion is about 300\,years.

{\it 12485$-$1543.}  The orbit of the 2.6 year inner pair Aa,Ab, first
computed by  \citet{Tok2014a}, is  updated using new  speckle measures
and  the unpublished  RVs of  both components  measured by  D.  Latham
(2012,  private communication).   The outer  component in  this nearby
triple  system,  at 2\farcs6  separation,  has  an  estimated period  of
400\,years.   \citet{Horch2017} published  recently  a similar  visual
orbit, apparently  unaware of the paper  by \citet{Tok2014a}. Compared
to Horch  et al., the coverage is  extended here by two  years and the
RVs are added, making the  new orbit  more accurate.  The
combined  orbit yields the  masses of  $0.96 \pm  0.08$ and  $0.75 \pm
0.07$ ${\cal M}_\odot$ and the orbital parallax of $36.7 \pm 1.1$ mas. 

{\it 13137$-$6248.}  The 25.6 year period of HDS~1852 is well defined
because it is now passing through the same part of the orbit where it
was discovered in 1991. Other orbital elements are still preliminary. 

{\it 14330$-$4224.} The pair of  A7V stars HDS 2054 passed through the
periastron  around  1998 without  being  observed  and  is now  slowly
opening. Its 66 year orbit  is preliminary, with a fixed eccentricity.
The mass  sum of 3.3 ${\cal  M}_\odot$ is less than  expected from the
spectral type.   If  the {\it Hipparcos}  measure is  flipped, the
  observed motion looks almost like a straight line, although its curvature
  is still  significant. A  rough circular orbit  with $P=222$  yr and
  $a=0\farcs33$ can be fitted to this arc.

{\it 15245$-$1322. } The preliminary  82 year orbit  of K7V  dwarfs appears
reasonably well constrained by the measures at SOAR and by those of \citet{Horch2017}.

{\it15339$-$1700.}    The   first   60 year   orbit   of   HDS~2185   by
\citet{Tok2012b} is updated here, as half of it is now covered.

{\it 17176+1025.} The orbit by \citet{Cve2013b} is revised to $P=46$ years. This 
chromospherically active star V2369~Oph is possibly young; it contains
an eclipsing subsystem with a period of 0\fd655. 

{\it  17195$-$5004.}   The  measures  of  FIN~356  with  nearly  equal
components,  allowing arbitrary  quadrant  changes, can  be fitted  by
several different orbits. However,  only the 12.9 year eccentric orbit
gives a reasonably  large mass sum of 2  ${\cal M}_\odot$. Its major
axis  is oriented  toward us,  the inclination  is 99\degr,  hence the
apparent separation is  always less than the true  semimajor axis. The
elements $e$ and $i$  are strongly correlated.  Interferometric and RV
monitoring  of the  next periastron  in  2023.6 is  needed. The  orbit
predicts  the   RV  difference   of  $\sim$30  km~s$^{-1}$   near  the
periastron.

{\it 17447$-$4244.} We reprocessed the observations of FIN~341 made at
SOAR in 2008 and 2009 to verify that the quadrant was opposite to that
of 2015--2017,  in agreement with the  proposed 15 year  orbit. One of
the two  archival speckle  measures made in  1989.3 fits the  orbit nicely,
while  the  separation  of  0\farcs15  measured  on  1991.39  strongly
contradicts both the orbit  (which predicts the separation of 20\,mas)
and  the non-resolution by  {\it Hipparcos}.   We discard  the 1991.39
speckle datum and use the  interferometric measures by W.  Finsen made
in 1959 and  1963.  This object belongs to  the $\lambda$ Bootis class
of   chemically  peculiar   stars   \citep{Paunzen1997},  making   the
measurement of  its mass particularly interesting.  The  masses of 1.9
and 1.5 ${\cal M}_\odot$ deduced from the orbit are slightly less than
expected for normal A2V stars.

{\it 18040+0150.} This is HD~165045, a double-lined spectroscopic
binary resolved at SOAR. The RVs measured by D. Latham (2012, private
communication) are used together with the speckle measures in the
combined orbit with $P=1.6$~year. Its small inclination prevents accurate 
measurement of the masses; the orbital parallax is 
$29.2 \pm 3.4$ mas.

{\it 18520$-$5418.}  The {\it  Hipparcos} astrometric binary was first
resolved at  Gemini in  2012 by \citet{Tok2013}  and then  followed at
SOAR.  The period  of this preliminary orbit is fixed  to 8 years. One
can't help noting  the discrepancy between the {\it  Gaia} parallax of
13.77\,mas, the HIP2 parallax of 19.5\,mas, and the dynamical parallax
of  21.5\,mas. The {\it  Gaia} parallax  leads to  the unrealistically
large mas sum of 6.2 ${\cal M}_\odot$ and is therefore suspect. With a
physical companion at 146\arcsec, the system is triple.

{\it 19377$-$4128.}  Both extremities of the 55 year circular orbit of
VOU~34 are covered by accurate speckle measures. The pair has made 1.5
revolutions since its first resolution in 1936. The moderate $\Delta V
\approx 0.4$ mag and the  opposite quadrants in 1991 ({\it Hipparcos})
and 2015 exclude the alternative eccentric orbit with half the
period. 

{\it 19598$-$0957.} The combined  5 year orbit by \citet{Pbx2000b} had
large residuals to the SOAR  measures, prompting its revision. The new
combined orbit uses  the published RVs and is  well defined (grade 2).
Its small  inclination and correspondingly small RV  amplitudes do not
help accurate measurement of the masses and orbital parallax, which is
$40.0 \pm 5.8$ mas.

{\it 20212$-$5147.} The {\it Hipparcos} binary HDS~2097 is quite fast,
with $P=10$~years and a well-defined first orbit. The star is on the
Magellan program of exo-planet search \citep{Arrigada2011}. 

{\it 21099$-$2424.}   The first 6.9  year orbit of HIP~104476  is well
defined  by  the  SOAR   data.   SIMBAD  calls  this  star  ``pre-main
sequence''; it is an X-ray  source.  The components are located on the
main sequence, however. 

{\it   21368$-$3043.}    The    preliminary   orbit   of   VOU~35   by
\citet{Tok2014a} is  revised here using recent  observations. The pair
has a  relatively close companion  C, discovered at SOAR  at 0\farcs45
from A and showing rapid  orbital motion (the estimated period of AB,C
is  $\sim$100  years). 


{\it 21400$-$5222.} The  SOAR data cover the same part  of the 21 year
orbit where  the pair was  resolved by {\it Hipparcos}  one revolution
earlier.   The new orbit  is preliminary.   The components  are nearly
equal;  both  are  located  on  the  subgiant branch  in  the  CMD  in
Figure~\ref{fig:cmd}, in agreement with the F7III spectral type.  

{\it  21504$-$5818.} The  first 17  year orbit  of HIP~107806  is well
covered. The  mass sum computed  with the {\it Hipparcos}  parallax of
24.1 mas  is 1.4  ${\cal M}_\odot$, appropriate  for the  G6V spectral
type. The dynamical parallax in Table~\ref{tab:ptm} is smaller and the
derived masses are correspondingly larger. 

{\it  22228$-$2937.}  HDS~3172 is a  fast pair  of G6V
dwarfs with $P=10.5$ years. Its first orbit (Figure~\ref{fig:orb}) is well
constrained.  The  resolution in  2016.9, at periastron  (38\,mas), is
tentative. Now the pair opens up again.

{\it 22535$-$1137.}  This is the bright triple system HR~8704 (74~Aqr)
where the primary  component of the interferometric pair  A,B (MCA 73)
is a 3.4 day  double-lined spectroscopic binary \citep{Catanzaro2004}.
The 19 year outer  orbit was first computed in  1993 and updated several
times since, the last one  by \citet{Msn2010c}. This is a typical case
where  the measurements  can be  modeled either  by a  highly inclined
orbit  with moderate  eccentricity  (the  19 year orbit)  or  by a  very
eccentric orbit with  half the period.  The ambiguity  can be resolved
when the true quadrants are known,  which is not the case for binaries
with  small $\Delta m$  and for  classical speckle  observations.  The
quadrants of MCA~73 with $\Delta  m \approx 0.6$ mag can be determined
from the speckle image reconstruction which is used at SOAR since 2015.  We
found  that in 2015  and 2016  the quadrants  differed from  the orbit
prediction,   whereas  they   matched  in   2008.76.    This  archival
observation  was reprocessed  to confirm  the quadrant;  see  also the
adaptive optics measurement  by \citet{Schoeller2010}.  Therefore, the
19 year  orbit  is  wrong, the  pair  is  always  resolved in  the  same
quadrant.   The   9.5 year  orbit   with  $e=0.86$  is   computed  here.
\citet{Catanzaro2004} give the RVs of  the center of  mass of  the close
pair (component A) and one  measurement of the secondary B.  These RVs
support  the eccentric  orbit (Figure~\ref{fig:MCA73})  which predicts
large RV variation near the  periastron.  The RV amplitudes are $K_1 =
10.0$  km~s$^{-1}$ and $K_2  = 19.6$  km~s$^{-1}$, the  systemic velocity  is $-2.7$
km~s$^{-1}$.  The RV amplitudes and inclination correspond to the mass sum of
$\sim$15   ${\cal  M}_\odot$;   however,  the   large  error   of  the
inclination, $i = 30\degr \pm 17\degr$, makes the spectroscopic masses
quite  uncertain.   The HIP2  parallax  of  3.95\,mas,  also not  very
precise, leads to  the mass sum of 16.6\,  ${\cal M}_\odot$. The orbit
thus  roughly matches  the  estimated  mass sum  of  three B9V  stars,
$\sim$11 ${\cal M}_\odot$.

This  chemically   peculiar  HgMn  star   has  attracted  considerable
attention   (135    references   in   SIMBAD).     Spectroscopic   and
interferometric coverage of the next  periastron in 2019.5 may lead to
the  more  accurate  measurements  of  the masses  and  the  distance,
especially  if the  closest approach  at 6\,mas  can be  resolved with
long-baseline  interferometers.  The   semimajor  axis  of  the  inner
spectroscopic pair is 0.3\,mas.

\section{Summary}
\label{sec:sum}

Speckle monitoring  of close visual  binaries, started at SOAR  in 2008,
begins  to bear  fruits  by defining   previously unknown  orbital
elements. Many southern binaries discovered by {\it Hipparcos} had not
been followed  and made one or more revolutions before their orbits
could be determined. Although some orbits are now well constrained,
there are still many binaries with sparse coverage where more
observations are needed. Their short (on the order of a decade) periods 
mean that this work can be accomplished in several years if the
speckle program at SOAR continues. 

We  feel that the  knowledge of  orbits will  be critical  for correct
interpretation of the {\it Gaia}  astrometry. As a reward, a large set
of  accurately measured  stellar masses  will become  available. Apart
from many common binaries composed of two normal dwarfs, there will be
interesting exceptions  such as massive  or evolved stars,  members of
clusters  and  hierarchical  multiple  systems,  stars  with  peculiar
spectra,   unusual  chemical  composition,   or  young   age.  Accurate
measurements of their masses will be of great value.


\acknowledgments 

 The  detailed report  by the Referee,  B.~Mason, who  checked the
  orbits and  assigned the  grades, helped to  improve this  paper. 
This work  used the  SIMBAD service operated  by Centre  des Donn\'ees
Stellaires  (Strasbourg, France),  bibliographic  references from  the
Astrophysics Data  System maintained  by SAO/NASA, and  the Washington
Double Star Catalog maintained at USNO.

{\it Facilities:}  \facility{SOAR}.

\clearpage


\LongTables

\begin{deluxetable}{c l rrr rrr r ccll}

\tabletypesize{\scriptsize}
\tablewidth{0pt}
\tablecaption{Orbital Elements \label{tab:orb}}
\tablehead{
\colhead{WDS} & 
\colhead{Discoverer} &
\colhead{$P$} & 
\colhead{$T_0$} &
\colhead{$e$} & 
\colhead{$a$} & 
\colhead{$\Omega$} &
\colhead{$\omega$} &
\colhead{$i$} &
\colhead{Gr} &
\colhead{Orbit} \\
 \colhead{HIP}   &
\colhead{Designation} &
\colhead{(yr)} & 
\colhead{(yr)} &
\colhead{ } & 
\colhead{($''$)} & 
\colhead{(\degr)} &
\colhead{(\degr)} &
\colhead{(\degr)} &
\colhead{ } & 
\colhead{Reference} 
}
\startdata
00135$-$3650 & HDS 32 & 15.34 & 2009.45 & 0.25 & 0.229 & 88.2 & 260.6 & 154.2 & 3 & New\tablenotemark{b} \\
  1083 &    & $\pm$0.35 & $\pm$0.73 & $\pm$0.08 & $\pm$0.016 & $\pm$48.5 & $\pm$38.4 & $\pm$11.3&     &  \\
00164$-$7024 & HEI 198 & 63.3 & 2010.7 & 0.42 & 0.166 & 153.2 & 210.0 & 121.6 & 5 & New\tablenotemark{b} \\
  1309 &    & $\pm$30.8 & $\pm$2.7 & $\pm$0.15 & $\pm$0.072 & $\pm$10.6 & $\pm$25.5 & $\pm$15.0&     &  \\
00258$+$1025 & HDS 57 & 48.87 & 2020.24 & 0.65 & 0.149 & 28.2 & 72.2 & 123.2 & 4 & New \\
  2035 &    & $\pm$7.32 & $\pm$0.60 & fixed & $\pm$0.015 & $\pm$8.1 & $\pm$3.7 & $\pm$1.9&     &  \\
01334$-$4354 & HDS 205 & 7.766 & 2011.025 & 0.493 & 0.1308 & 198.0 & 115.5 & 65.5 & 2 & New\tablenotemark{a} \\
  7254 &    & $\pm$0.041 & $\pm$0.014 & $\pm$0.012 & $\pm$0.0013 & $\pm$0.8 & $\pm$1.1 & $\pm$0.6&     &  \\
02166$-$5026 & TOK 185 & 13.97 & 2013.89 & 0.20 & 0.1038 & 276.8 & 15.8 & 50.0 & 4 & Tok2015c \\
 10611 &    & $\pm$4.44 & $\pm$0.25 & $\pm$0.19 & $\pm$0.0247 & $\pm$8.0 & $\pm$10.4 & $\pm$7.2&     &  \\
02390$-$5811 & HDS 345 & 6.019 & 2016.894 & 0.516 & 0.1976 & 257.5 & 94.5 & 138.1 & 3 & New\tablenotemark{b} \\
 12351 &    & $\pm$0.020 & $\pm$0.007 & $\pm$0.011 & $\pm$0.0026 & $\pm$3.2 & $\pm$2.5 & $\pm$1.3&     &  \\
04007$+$2023 & TOK 363 Aa,Ab & 16.761 & 2010.259 & 0.311 & 0.1141 & 198.2 & 146.5 & 31.5 & 5(3) & Gri2012\tablenotemark{a,c}\\
 18719 &    & $\pm$0.115 & $\pm$0.191 & $\pm$0.014 & $\pm$0.0030 & $\pm$3.0 & $\pm$4.2 & $\pm$4.4&     &  \\
04049$-$3527 & CHR 224 BC & 52.17 & 2009.67 & 0.920 & 0.2714 & 85.4 & 153.6 & 123.2 & 4 & New\tablenotemark{a} \\
 19052 &    & $\pm$2.35 & $\pm$0.07 & $\pm$0.006 & $\pm$0.0042 & $\pm$1.9 & $\pm$3.0 & $\pm$1.5&     &  \\
04074$-$6413 & LMP 51 & 26.028 & 1996.458 & 0.57  & 0.4201 & 294.8 & 221.0 & 96.5 & 5(4) & Jen2015 \\
 19233 &    & $\pm$0.076 & fixed & fixed & $\pm$0.0067 & $\pm$1.1 & fixed & $\pm$0.5&     &  \\
04107$-$0452 & A 2801 & 20.621 & 2014.031 & 0.8400 & 0.1672 & 153.3 & 68.6 & 69.0 & 2(1) &  Tok2014a,Gri2015\tablenotemark{c} \\
 19508 &    & $\pm$0.005 & $\pm$0.003 & $\pm$0.0014 & $\pm$0.0021 & $\pm$0.6 & $\pm$0.3 & $\pm$0.5&     &  \\
04422$+$0259 & A 2424 & 58.1 & 2011.05 & 0.345 & 0.1545 & 228.6 & 265.1 & 85.3 & 3 & Tok2015c \\
 21880 &    & $\pm$4.8 & $\pm$0.63 & $\pm$0.026 & $\pm$0.0085 & $\pm$0.8 & $\pm$8.3 & $\pm$0.6&     &  \\
05086$-$1810 & WSI 72 & 0.962 & 2016.174 & 0.258 & 0.0994 & 39.8 & 163.2 & 93.8 & 2 & New\tablenotemark{a} \\
 23932 &    & $\pm$0.002 & $\pm$0.016 & $\pm$0.025 & $\pm$0.0022 & $\pm$0.4 & $\pm$4.8 & $\pm$0.7&     &  \\
05103$-$0736 & A 484 & 18.90 & 1999.74 & 0.817 & 0.1719 & 290.3 & 123.2 & 104.2 & 3 & Hrt2012a \\
 24076 &    & $\pm$0.07 & $\pm$0.39 & $\pm$0.031 & $\pm$0.0183 & $\pm$1.2 & $\pm$6.2 & $\pm$2.2&     &  \\
05525$-$0217 & HDS 787 & 11.963 & 1999.710 & 0.244 & 0.1207 & 331.1 & 271.8 & 57.2 & 2 & Tok2014a \\
 27758 &    & $\pm$0.036 & $\pm$0.045 & $\pm$0.005 & $\pm$0.0007 & $\pm$0.6 & $\pm$0.7 & $\pm$0.5&     &  \\
06454$-$3148 & EHR 9 Ba,Bb & 6.909 & 2014.458 & 0.206 & 0.1173 & 193.6 & 349.5 & 140.6 & 3 & Tok2015c \\
 32366B&    & $\pm$0.051 & $\pm$0.066 & $\pm$0.016 & $\pm$0.0032 & $\pm$4.1 & $\pm$4.5 & $\pm$2.6&     &  \\
06533$-$1902 & CHR 169 & 36.28 & 2017.64 & 0.533 & 0.1928 & 353.7 & 257.8 & 105.3 & 3 & New \\
 33077 &    & $\pm$0.82 & $\pm$0.12 & $\pm$0.033 & $\pm$0.0049 & $\pm$1.0 & $\pm$1.5 & $\pm$1.3&     &  \\
07269$+$2015 & CHR 26 & 8.571 & 2000.494 & 0.412 & 0.0518 & 341.4 & 246.7 & 50.1 & 2 & Ole1998c \\
 36156 &    & $\pm$0.068 & $\pm$0.113 & $\pm$0.068 & $\pm$0.0025 & $\pm$5.7 & $\pm$6.7 & $\pm$5.0&     &  \\
08122$+$1739 & HUT 1 Ca,Cb & 17.263 & 1997.743 & 0.180 & 0.3592 & 81.0 & 287.3 & 150.0 & 4(3) & RAO2015,Gri2000\tablenotemark{c} \\
 40167 &    & $\pm$0.032 & $\pm$0.160 & $\pm$0.013 & $\pm$0.0058 & $\pm$1.9 & $\pm$3.6 & fixed   &     &  \\
08403$+$1921 & CHR 130 & 35.45  & 1987.86  & 0.169 & 0.1402 & 339.8 & 226.3 & 90.4 & 3 & Ole2002b \\
 42542 &    & $\pm$0.22  & $\pm$0.53  & $\pm$0.013 & $\pm$0.0021 & $\pm$0.6 & $\pm$5.8  & $\pm$0.6&     &  \\
08447$-$4238 & CHR 238 & 2.262 & 2012.584 & 0.676 & 0.0775 & 13.0 & 125.2 & 152.2 & 4 & Tok2015c \\
 42916 &    & $\pm$0.001 & $\pm$0.007 & $\pm$0.007 & $\pm$0.0015 & $\pm$5.2 & $\pm$5.1 & $\pm$3.8&     &  \\
09252$-$1258 & WSI 73 & 13.15 & 2018.94 & 0.61 & 0.1443 & 93.6 & 52.2 & 87.0 & 3 & Tok2015a \\
  46191 &    & $\pm$0.61 & $\pm$1.40 & fixed & $\pm$0.0304 & $\pm$1.5 & $\pm$19.7 & $\pm$1.5&     &  \\
10121$-$0241 & DEL 3 & 6.365 & 2015.502 & 0.50 & 0.2624 & 249.7 & 94.7 & 91.8 & 4 & New \\
 49969 &    & $\pm$0.027 & $\pm$0.030 & fixed & $\pm$0.0051 & $\pm$0.6 & $\pm$0.8 & $\pm$0.4&     &  \\
10294$+$1211 & HDS 1507 & 15.59 & 2011.79 & 0.372 & 0.0984 & 277.0 & 286.3 & 24.6 & 3 & Cve2016b\tablenotemark{b} \\
 51360 &    & $\pm$0.11 & $\pm$0.36 & $\pm$0.032 & $\pm$0.0021 & $\pm$29.7 & $\pm$24.7 & $\pm$6.9&     &  \\
10529$-$1717 & HDS 1556 & 14.95 & 2003.60 & 0.553 & 0.1875 & 109.3 & 61.8 & 97.0 & 2 & Tok2015c \\
 53206 &    & $\pm$0.26 & $\pm$0.22 & $\pm$0.032 & $\pm$0.0056 & $\pm$0.3 & $\pm$2.3 & $\pm$0.5&     &  \\
11420$-$1701 & TOK 384 Aa,Ab & 21.1 & 2016.21 & 0.58 & 0.163 & 169.0 & 259.6 & 21.1 & 5 & New\tablenotemark{b} \\
  57078 &    & $\pm$8.7 & $\pm$0.10 & $\pm$0.12 & $\pm$0.040 & $\pm$40.5 & $\pm$46.5 & $\pm$5.4&     &  \\
12485$-$1543 & WSI 74  Aa,Ab & 2.661 & 2011.094 & 0.508 & 0.0843 & 146.6 & 145.6 & 55.3 & 2 & Tok2014a\tablenotemark{c} \\
 62505 &    & $\pm$0.002 & $\pm$0.005 & $\pm$0.007 & $\pm$0.0008 & $\pm$0.9 & $\pm$1.3 & $\pm$0.8&     &  \\
13137$-$6248 & HDS 1852 & 25.644 & 1996.680 & 0.291 & 0.1659 & 149.1 & 124.0 & 63.1 & 5 & New\tablenotemark{b} \\
 64537 &    & $\pm$0.401 & $\pm$1.306 & $\pm$0.068 & $\pm$0.0059 & $\pm$3.3 & $\pm$15.0 & $\pm$1.9&     &  \\
14330$-$4224 & HDS 2054 & 66.5 & 1997.84 & 0.80  & 0.207 & 356.7 & 116.0 & 97.5 & 4 & New \\
  71140 &    & $\pm$22.2& $\pm$2.36 & fixed & $\pm$0.049 & $\pm$2.2 & $\pm$6.0 & $\pm$1.6&     &  \\
15245$-$1322 & HDS 2167 & 82.30 & 2008.37 & 0.422 & 0.440 & 303.5 & 157.1 & 108.9 & 4 & New \\
 75416 &    & $\pm$10.11 & $\pm$0.34 & $\pm$0.049 & $\pm$0.037 & $\pm$0.8 & $\pm$2.4 & $\pm$0.6&     &  \\
15339$-$1700 & HDS 2185 & 72.32 & 2010.26 & 0.414 & 0.5757 & 150.6 & 310.3 & 34.6 & 4 & Tok2012b \\
  76203 &    & $\pm$2.20 & $\pm$0.10 & $\pm$0.012 & $\pm$0.0069 & $\pm$1.1 & $\pm$1.1 & $\pm$1.0&     &  \\
17176$+$1025 & HDS 2445 & 45.80 & 2009.14 & 0.149 & 0.2781 & 84.7 & 140.4 & 97.2 & 3 & Cve2013b\tablenotemark{b} \\
 84595 &    & $\pm$3.44 & $\pm$2.70 & $\pm$0.070 & $\pm$0.0063 & $\pm$0.9 & $\pm$23.7 & $\pm$0.5&     &  \\
17195$-$5004 & FIN 356 & 12.93 & 2010.65 & 0.814 & 0.114 & 85.9 & 93.6 & 98.6 & 2 & New \\
 84759 &    & $\pm$0.15 & $\pm$0.16 & $\pm$0.053 & $\pm$0.016 & $\pm$1.0 & $\pm$1.4 & $\pm$1.9&     &  \\
17447$-$4244 & FIN 341 & 15.12 & 2012.71 & 0.165 & 0.1183 & 179.5 & 244.6 & 81.7 & 5 & New \\
 86847 &    & $\pm$0.10 & $\pm$0.68 & $\pm$0.038 & $\pm$0.0049 & $\pm$0.8 & $\pm$14.4 & $\pm$1.6&     &  \\
18040$+$0150 & TOK 695 & 1.615 & 1996.698 & 0.501 & 0.0472 & 271.0 & 287.4 & 37.7 & 4(3) & New\tablenotemark{c} \\
 88481 &    & $\pm$0.001 & $\pm$0.005 & $\pm$0.011 & $\pm$0.0016 & $\pm$2.5 & $\pm$2.5 & $\pm$3.8&     &  \\
18520$-$5418 & TOK 325 Aa,Ab & 8.0 & 2017.985 & 0.413 & 0.1012 & 92.4 & 41.2 & 60.8 & 4 & New \\
 92592 &    & fixed & $\pm$0.136 & $\pm$0.029 & $\pm$0.0029 & $\pm$2.5 & $\pm$4.6 & $\pm$2.0&     &  \\
19377$-$4128 & VOU 34 & 55.0 & 2003.75 & 0.02 & 0.1642 & 134.4 & 86.0 & 96.9 & 3 & New \\
 96545 &    & $\pm$6.8 & $\pm$4.40 & $\pm$0.12 & $\pm$0.0045 & $\pm$0.6 & $\pm$29.4 & $\pm$1.1&     &  \\
19598$-$0957 & HO 276 & 4.867 & 2012.022 & 0.605 & 0.1514 & 326.6 & 319.4 & 18.8 & 2 & Pbx2000b\tablenotemark{c} \\
 98416 &    & $\pm$0.004 & $\pm$0.011 & $\pm$0.004 & $\pm$0.0013 & $\pm$1.8 & $\pm$2.2 & $\pm$2.7&     &  \\
20212$-$5147 & HDS 2907 & 9.886 & 2017.514 & 0.651 & 0.1948 & 32.7 & 194.0 & 48.8 & 3 & New \\
100356 &    & $\pm$0.129 & $\pm$0.016 & $\pm$0.016 & $\pm$0.0047 & $\pm$3.2 & $\pm$5.5 & $\pm$2.3&     &  \\
21099$-$2424 & HDS 3015 & 6.88 & 2014.99 & 0.566 & 0.0882 & 296.3 & 13.7 & 66.0 & 2 & New\tablenotemark{a} \\
104476 &    & $\pm$0.12 & $\pm$0.18 & $\pm$0.032 & $\pm$0.0040 & $\pm$4.5 & $\pm$12.2 & $\pm$2.0&     &  \\
21368$-$3043 & VOU 35 & 19.547 & 2018.160 & 0.390 & 0.1673 & 126.9 & 91.4 & 107.8 & 3 & Tok2014a \\
106701 &    & $\pm$0.086 & $\pm$0.088 & $\pm$0.033 & $\pm$0.0037 & $\pm$1.2 & $\pm$1.4 & $\pm$0.8 &     &  \\
21400$-$5222 & HDS 3084 & 21.31 & 1998.80 & 0.54 & 0.0762 & 31.5 & 231.4 & 150.4 & 3 & New\tablenotemark{b} \\
106978 &    & $\pm$0.74 & $\pm$1.45 & $\pm$0.18 & $\pm$0.0084 & $\pm$74.2 & $\pm$66.6 & $\pm$11.1&     &  \\
21504$-$5818 & HDS 3109 & 16.79 & 2004.15 & 0.246 & 0.1782 & 307.5 & 255.3 & 86.9 & 3 & New \\
107806 &    & $\pm$0.75 & $\pm$0.46 & $\pm$0.062 & $\pm$0.0039 & $\pm$0.4 & $\pm$3.7 & $\pm$0.5&     &  \\
22228$-$2937 & HDS 3172 &10.51  & 2017.374 & 0.288 & 0.1444 & 155.4 & 325.7 &  97.4 & 2 & New\tablenotemark{a} \\
110483 &    & $\pm$0.25  & $\pm$0.069 & $\pm$0.023 & $\pm$0.0031 & $\pm$0.4 & $\pm$3.1 & $\pm$0.5&     &  \\
22535$-$1137 & MCA 73 & 9.479 & 2010.039 & 0.862 & 0.0460 & 40.9 & 70.7 & 29.8 & 2 & Msn2010c \\
113031 &    & $\pm$0.044 & $\pm$0.134 &$\pm$0.029  & $\pm$0.0061 & $\pm$20.3 & $\pm$16.1 & $\pm$17.4&     &  
\enddata
\tablenotetext{a}{See Figure~1.}
\tablenotetext{b}{Insufficient coverage.}
\tablenotetext{c}{Combined orbit using the RVs.}
\tablerefs{
Cve2013b -- \citet{Cve2013b};
Cve2016b -- \citet{Cve2016b};
Jen2015 -- \citet{Jen2015};
Gri2000 -- \citet{Griffin2000};
Gri2012 -- \citet{Griffin2012};  
Gri2015 -- \citet{Gri2015};  
Hor2017 -- \citet{Horch2017};
Hrt2012a -- \citet{Hrt2012a};  
Msn2010c -- \citet{Msn2010c};
Ole1998c -- \citet{Ole1998c};
Ole2002b -- \citet{Ole2002b};
Pbx2000b --  \citet{Pbx2000b}; 
RAO2015 -- \citet{RAO2015};
Tok2012b -- \citet{Tok2012b};
Tok2014a -- \citet{Tok2014a}; 
Tok2015c -- \citet{Tok2015c}.
}
\end{deluxetable}

\begin{deluxetable}{c r rrr rr r}
\tabletypesize{\scriptsize}
\tablewidth{0pt}
\tablecaption{Observations and residuals (Fragment) \label{tab:obs}}
\tablehead{
\colhead{WDS} & 
\colhead{$T$} &
\colhead{$\theta$} & 
\colhead{$\rho$} &
\colhead{$\sigma$} & 
\colhead{O$-$C$_\theta$} & 
\colhead{O$-$C$_\rho$} &
\colhead{Flag} \\
& \colhead{(yr)} & 
\colhead{(\degr)} &
\colhead{($''$)} & 
\colhead{($''$)} & 
\colhead{(\degr)} &
\colhead{($''$)} &
}
\startdata
00135$-$3650 &  1991.2500 &  282.0 &   0.2190 &   0.0100 &    0.0 & $-$0.0000 & H  \\
00135$-$3650 &  2014.7662 &   46.3 &   0.2565 &   0.0030 &   $-$0.0 &  $-$0.0000 & S \\
00135$-$3650 &  2015.9130 &   28.0 &   0.2600 &   0.0030 &    0.1 &  $-$0.0000 & S \\
00135$-$3650 &  2016.9506 &   11.3 &   0.2592 &   0.0030 &   $-$0.0 &   0.0001 & S 
\enddata
\end{deluxetable}

\clearpage

\begin{deluxetable}{ccc cc cc rr rr rr}
\tabletypesize{\scriptsize}
\tablewidth{0pt}
\tablecaption{Parallaxes and photometry \label{tab:ptm}}
\tablehead{
\colhead{WDS} &
\colhead{HIP} &
\colhead{Spectral} & 
\colhead{${\pi_{\rm HIP2}}$} & 
\colhead{${\pi_{\rm dyn}}$} & 
\colhead{${\cal M}_1$} & 
\colhead{${\cal M}_2$} & 
\colhead{$V$} &
\colhead{$V-I_C$} &
\colhead{$\Delta y$} &
\colhead{$\sigma_{\Delta y}$} &
\colhead{$\Delta I$} &
\colhead{$\sigma_{\Delta I}$} \\
&  & 
\colhead{Type} &  
\colhead{(mas)} & 
\colhead{(mas)} & 
\colhead{(${\cal M}_\odot$)} & 
\colhead{(${\cal M}_\odot$)} & 
\colhead{(mag)} & 
\colhead{(mag)} & 
\colhead{(mag)} & 
\colhead{(mag)} & 
\colhead{(mag)} & 
\colhead{(mag)} 
}
\startdata
00135$-$3650 &   1083 & M0.5V &   35.5 &   36.0*  &     0.59 &     0.52 &    10.77 &     1.78 & \ldots & \ldots &     0.77 &     0.06 \\
00164$-$7024 &   1309 & F0V &    5.2 &    6.9  &     1.84 &     1.61 &     7.18 &     0.28 &     0.68 & \ldots &     0.55 &     0.02 \\
00258$+$1025 &   2035 & G0 &    9.1 &    8.7  &     1.19 &     0.93 &     9.21 &     0.71 & \ldots & \ldots &     1.04 &     0.21 \\
01334$-$4354 &   7254 & K0/K1V &   26.3 &   27.5* &     0.97 &     0.81 &     7.83 &     0.86 &     1.21 &     0.18 &     0.88 &     0.04 \\
02166$-$5026 &  10611 & G5V &   16.9 &   14.7  &     1.00 &     0.82 &     9.04 &     0.75 &     1.22 &     0.74 &     1.19 &     0.52 \\
02390$-$5811 &  12351 & M0Ve &   59.4 &   57.0*  &     0.60 &     0.55 &     9.48 &     1.52 & \ldots & \ldots &     0.45 &     0.03 \\
04007$+$2023 &  18719 & G4V &   16.1 &   14.7* &     1.12 &     0.53 &     8.66 &     0.75 &     5.17 & \ldots &     2.36 &     0.73 \\
04049$-$3527 &  19052B&\ldots & 14.0 &   16.4 &     0.97 &     0.71 &     9.14 &     0.70 &     2.06 &     0.41 &     1.47 &     0.09 \\
04074$-$6413 &  19233 & G3V &   42.3 &   40.3  &     1.14 &     0.53 &     6.37 &     0.70 & \ldots & \ldots &     3.79 &     0.68 \\
04107$-$0452 &  19508 & G3/5IV &   16.1 &   16.8* &     1.24 &     1.08 &     7.37 &     0.71 &     0.80 &     0.07 &     0.55 &     0.31 \\
04422$+$0259 &  21880 & A0 &    5.7 &    6.8* &     1.83 &     1.58 &     7.26 &     0.21 &     0.74 &     0.56 &     0.28 &     0.16 \\
05086$-$1810 &  23932 & M5 &  107.8 &  118.5* &     0.34 &     0.29 &    10.28 &     2.64 &     0.40 &     0.56 &     0.24 &     0.15 \\
05103$-$0736 &  24076 & F8 &   17.4 &   18.6* &     1.15 &     1.08 &     7.42 &     0.62 &     0.35 &     0.14 &     0.20 &     0.02 \\
05525$-$0217 &  27758 & F5 &   20.1 &   17.6* &     1.29 &     0.97 &     7.26 &     0.63 &     1.62 &     0.28 &     1.33 &     0.08 \\
06454$-$3148 &  32366B& M3? &   39.6 &  \ldots &     0.25 &     0.15 &    13.60 &     2.70 & \ldots & \ldots &     1.17 &     1.25 \\
06533$-$1902 &  33077 & A9/F0III &   10.5 &   11.7*  &     2.17 &     1.22 &     5.65 &     0.32 &     2.95 &     0.20 & \ldots & \ldots \\
07269$+$2015 &  36156 & F2Vn &    7.3 &    7.6* &     2.17 &     2.13 &     5.94 &     0.39 &     0.09 &     0.15 & \ldots & \ldots \\
08122$+$1739 &  40167 & G0V &   39.9 &   40.8  &     1.54: &     0.75: &     4.67 &     0.60 &     4.33 & \ldots &     2.66 & \ldots \\
08403$+$1921 &  42542 & A9V &    7.2 &    8.8  &     1.91 &     1.35 &     6.76 &     0.32 &     1.78 &     0.17 & \ldots & \ldots \\
08447$-$4238 &  42916 & G5V &   39.6 &   37.5  &     0.99 &     0.74 &     7.20 &     0.81 &     1.88 &     0.47 &     1.14 & \ldots \\
09252$-$1258 &  46191 & K1/2V &   23.0 &   22.9* &     0.78 &     0.66 &     9.62 &     1.00 &     1.17 &     0.39 &     0.38 &     0.20 \\
10121$-$0241 &  49969 & M0 &   81.1 &   86.3  &     0.44 &     0.25 &    10.64 &     2.32 & \ldots & \ldots &     1.00 &     0.05 \\
10294$+$1211 &  51360 & F2 &   11.0 &   11.3*  &     1.53 &     1.18 &     7.25 &     0.51 &     1.43 &     0.08 &     1.21 &     0.15 \\
10529$-$1717 &  53206 & G6V &   25.5 &   24.0* &     1.10 &     1.03 &     7.13 &     0.74 &     0.36 &     0.17 &     0.25 & \ldots \\
11420$-$1701 &  57078 & F5V &   17.0 &   17.2  &     1.27 &     0.67 &     7.61 &     0.56 &     3.98 &     0.54 &     2.86 &     0.31 \\
12485$-$1543 &  62505 & K2V &   42.0 &   37.8* &     0.87 &     0.70 &     7.93 &     0.99 &     1.43 &     0.15 &     1.07 &     0.13 \\
13137$-$6248 &  64537 & G5 &   15.7 &   15.8  &     0.94 &     0.83 &     9.14 &     0.83 &     0.76 &     0.21 &     0.31 &     0.18 \\
14330$-$4224 &  71140 & A7/8IV &    7.8 &    9.0  &     1.45 &     1.31 &     7.78 &     0.32 &     0.56 &     0.32 &     0.32 &     0.33 \\
15245$-$1322 &  75416 & K7V &   24.1 &   20.9  &     0.71 &     0.67 &    10.27 &     1.26 & \ldots & \ldots &     0.31 &     0.16 \\
15339$-$1700 &  76203 & G9IV-V &   25.9 &   28.0 &     0.94 &     0.72 &     8.14 &     0.88 &     1.78 &     0.06 &     1.36 & \ldots \\
17176$+$1025 &  84595 & G5 &   14.4 &   17.7* &     1.04 &     0.81 &     8.47 &     0.86 & \ldots & \ldots &     1.10 & \ldots \\
17195$-$5004 &  84759 & F2III &   16.9 &   14.2* &     1.56 &     1.50 &     6.27 &     0.48 &     0.23 &     0.22 & \ldots & \ldots \\
17447$-$4244 &  86847 & A2/A3IV/V &   13.9 &   12.9 &     1.89 &     1.48 &     5.87 &     0.18 &     1.24 &     0.08 & \ldots & \ldots \\
18040$+$0150 &  88481 & G5 &   29.0 &   29.0 &     0.92 &     0.74 &     8.14 &     0.83 &     1.37 &     0.14 &     1.00 &     0.07 \\
18520$-$5418 &  92592 & G3V &   19.5 &   21.5  &     0.98 &     0.65 &     8.54 &     0.69 &     2.81 &     0.18 &     1.75 &     0.43 \\
19377$-$4128 &  96545 & F3IV &    6.8 &    7.9* &     1.56 &     1.43 &     7.65 &     0.51 &     0.46 &     0.05 &     0.25 &     0.10 \\
19598$-$0957 &  98416 & F5V &   45.0 &   41.2* &     1.18 &     0.91 &     5.88 &     0.67 &     1.57 &     0.04 &     1.46 & \ldots \\
20212$-$5147 & 100356 & K7 &   41.7 &   40.4* &     0.59 &     0.56 &    10.23 &     1.73 & \ldots & \ldots &     0.34 &     0.18 \\
21099$-$2424 & 104476 & G5V &   19.1 &   19.5*  &     1.03 &     0.92 &     8.09 &     0.79 &     0.68 &     0.28 &     0.32 &     0.15 \\
21368$-$3043 & 106701 & G5V &   16.1 &   19.0*  &     0.94 &     0.86 &     8.67 &     0.76 &     0.58 &     0.62 &     0.34 &     0.42 \\
21400$-$5222 & 106978 & F7III &    5.8 &    6.0* &     2.34 &     2.20 &     6.20 &     0.67 &     0.27 &     0.19 &     0.32 & \ldots \\
21504$-$5818 & 107806 & G6V &   24.1 &   21.8* &     1.00 &     0.94 &     7.89 &     0.78 &     0.39 &     0.17 &     0.20 &     0.13 \\
22228$-$2937 & 110483 & G6V &   25.5 &   25.5* &     0.98 &     0.67 &     8.19 &     0.80 &     2.50 &     0.16 &     1.92 &     0.16 \\
22535$-$1137 & 113031 & B8IV/V &    4.0 &    6.0*  &     2.62 &     2.29 &     5.80 & \ldots &     0.62 &     0.13 &     0.58 & \ldots 
\enddata
\end{deluxetable}

\end{document}